# On the Bound of Cumulative Return in Trading Series and the Verification Using Technical Trading Rules

Can Yang*¶, Junjie Zhai¶, Helong Li*

School of Computer Science and Engineering, South China University of Technology, Guangzhou, Guangdong, China

*Corresponding author, E-mail: cscyang@scut.edu.cn(CY), hlongli@scut.edu.cn

¶These authors contributed equally to this work.

## Abstract

Although there is a wide use of technical trading rules in stock markets, the profitability of them still remains controversial. This paper first presents and proves the upper bound of cumulative return, and then introduces many of conventional technical trading rules. Furthermore, with the help of bootstrap methodology, we investigate the profitability of technical trading rules on different international stock markets, including developed markets and emerging markets. At last, the results show that the technical trading rules are hard to beat the market, and even less profitable than the random trading strategy.

## I.  Introduction

As a significant part of financial markets, stock market undoubtedly becomes the focus of many analysts and investors. However, stock price is always affected by political, macro or micro economy events, and it is a challenge to make profit using an existing trading rule. Barak et al. [1] indicated that stock market is a complex dynamic system and it is difficult to make a forecast. While driven by high profits, lots of investors still involve in stock investment, including professional and unprofessional ones.

For professional investors, fundamental analysis and technical analysis are two major approaches in making decisions in stock markets [2]. Fundamental analysis is based on expectations about future asset's prices upon market fundamentals and economic factors, including macroeconomic, industrial and business variables. Whereas, technical analyses extrapolate the trend or statistically relevant characteristics from past stock prices, on the base of the assumption that historical behavior has an effect on the future stock price. The empirical evidences have shown that both fundamental and technical analysis can help to achieve great profits in stock market investment [3]. However, for unprofessional individual investors, lack of generous financial advantage and information superiority always makes them become victims when institutional investors manipulate stock price [7]. Unwilling to accept such a fact, some of them try to seek help from technical trading rules in investment (since a qualitative analysis on macroeconomics fundamentals is usually subjective and hard to assess [8]). But, can simple technical trading rules help them to achieve steady profits? This article will give out our response.

As for previous literature, in the 1960s and 1970s, most of studies supported random walk hypothesis and claimed that technical analysis is invalid [9]. Therefore, technical analysis was largely dismissed by academics [11]. Nevertheless, in 1992, Brock et al. [12] tested the most popular technical trading rules by using the Dow Jones Industrial Average during the period from 1897 to 1986 and found their significant predictive ability. Their work totally changed the attitude of academic view on technical analysis. On the basis of their work, many analysts found that technical trading rules are effective in forecasting the market trend and making profits [2]. In spite of its popularity in practitioners, some scholars still kept a skeptical attitude and provided the evidence to question its effectiveness. Pierre and Olivier [16] used a new approach called False Discovery Rate (FDR) and persistence test to verify the merits of technical trading rules, and their results seriously call into question the economic value of technical trading strategies. Similarly, Biondo et al. [8] investigated the predictability of several most used technical trading rules by comparing with the random trading strategy in FTSE-UK, FTSE-MIB, DAX, and S&P 500 indexes, and their results indicated the predictability of technical trading strategies could not beat the random trading strategy. Actually, the profitability of technical trading strategies is also questioned. After performing a true out-of-sample test in contrast to Brock et al.'s [12], Fang et al. [11] made an opposite conclusion that simple technical trading rules have no significant profitability. Moreover, Zhu et al. [17] investigated the profitability of moving average (MA) and trading range break (TRB) rules in Chinese stock markets, and found that simple trading rules like MA and TRB could not beat the standard buy-and-hold strategy and once transaction costs are considered, trading profits will be eliminated completely. However, emerging markets usually have different behaviors in contrast to developed markets [18], because emerging markets always dominated by less experienced individual investors [19], while developed markets are dominated by sophisticated institutional investors [21]. Therefore, it is necessary to examine whether the profitability of technical trading strategies differs between emerging markets and developed markets.

In this paper, we first present and prove the upper bound of cumulative return, and then introduce many conventional technical trading rules. In addition, based on bootstrap methodology, we investigate the profitability of these technical trading rules on both emerging market and developed market. In a word, our contribution can be summarized as follows:

1. Presentation and Proof on the upper bound of cumulative return in theory.
2. The finding that when the mean of return rate series ($\bar{r}$) is less than the transaction cost rate ($k$), the more trades, the more losses.
3. The finding the profitability of the conventional technical trading rules is not stable, and they rarely beat the market or even the random trading strategy.

The rest of this paper is organized as follows. Section II first present and prove the bound of cumulative return, In Section III, we introduce the used dataset, simple technical trading rules, and the testing method of bootstrap in this work. And we explore the profitability of these technical trading rules on distinct stock indices in Section IV. We conclude in the end.

# II. On the Bound of Cumulative Return

Assume $S$ is a time series. The return of the i-th trade is defined as,

$$r_i = \frac{S(s_i) - S(b_i)}{S(b_i)} \tag{1}$$

Here, $S(*)$ is the price at time $*$. $s_i$ and $b_i$ denotes the time of conducting i-th buy and sell trades, respectively. When take transaction costs into consideration, the cumulative return $R(n)$ can be calculated by:

$$R(n) = \prod_{i=1}^{n}\left(1 + \frac{S(s_i) - S(b_i) - T_i}{S(b_i)}\right) \tag{2}$$

Where n denotes the number of trades, and $T_i$ denotes the transaction costs in the i-th trade, which can be approximately considered as $T_i \approx S(s_i) \cdot k$, and here $k$ is regarded as a constant, denoting transaction cost rate. Further, we can obtain:

$$R(n) = \prod_{i=1}^{n}(1-k)\frac{S(s_i)}{S(b_i)} = \prod_{i=1}^{n}(1-k)(1+r_i) \tag{3}$$

Before deriving the bounds of cumulative return $R(n)$, we first introduce D-I inequality which is proposed by Dragomir and Ionescu[23]. As a preliminary knowledge, we show the D-I inequality as follows:

**Lemma 1 (D-I inequality)** *Let $f: I \subseteq R \to R$ be a differentiable convex function on $I$, $x_i \in I$ and $p_i \geq 0 (i = 1, ..., n)$ with $P_n := \sum_{i=1}^{n} P_i > 0$. Then we have the inequality as follow,*

$$0 \leq \frac{1}{P_n}\sum_{i=1}^{n} p_i f(x_i) - f\left(\frac{1}{P_n}\sum_{i=1}^{n} p_i x_i\right) \tag{4}$$

*Where $f'(x_i)$ is the first derivation of $f(x)$ at $x_i$, and $f'(x_i) = \frac{\partial f(x_i)}{\partial x}$. The proof on **Lemma 1** can be found in [23].*

Then, based on the D-I inequality, we exhibit the upper bound of cumulative return $R(n)$ as Theorem 1:

**Theorem 1 (Upper bound)** $\forall n > 0, \exists R(n)$ *satisfies the following inequality:*

$$R(n) \leq [(1-k)(1+r_i)]^n \tag{5}$$

Here $\bar{r} = \frac{1}{n}\sum_{i=1}^{n} r_i$, $k$ is transaction cost rate.

**Proof :** *Let $G(n)$ be $\ln R(n)$, according to Eq. 3, then*

$$G(n) = -\ln \prod_{i=1}^{n}(1-k)(1+r_i) = n \cdot [-\ln(1-k)] + \sum_{i=1}^{n}[-\ln(1+r_i)] \tag{6}$$

*In the D-I inequality, we let $p_i = 1$, so $P_n := \sum_{i=1}^{n} P_i = n$, and $f(x) = -\ln x$. It is easily to certify that $f(x)$ is a differentiable convex function on $(0, +\infty)$, because the second derivative of $f(x)$ is greater than 0 when $x \in (0, +\infty)$ (see Eq. 7).*

$$\frac{\partial^2 f}{\partial x^2} = \frac{\partial(\partial(-\ln x)/\partial x)}{\partial x} = \frac{\partial(-1/x)}{\partial x} = \frac{1}{x^2} > 0 \tag{7}$$

*Let $x_i = 1 + r_i (-1 \leq r_i)$, the Eq. 4 can be transformed into:*

$$-n \cdot \ln\left[\frac{1}{n}\sum_{i=1}^{n}(1+r_i)\right] \leq \sum_{i=1}^{n}[-\ln(1+r_i)] \tag{8}$$

*With the assistance of Eq. 8 and the Eq. 6, $G(n)$ can be further derived:*

$$G(n) \geq n \cdot [-\ln(1-k)] - n \cdot \ln(1+\bar{r}) = -\ln[(1-k)(1+r_i)]^n \tag{9}$$

*Where $\bar{r} = \frac{1}{n}\sum_{i=1}^{n} r_i$, and considering that $G(n) = -\ln R(n)$, we can further obtain*

$$R(n) \leq [(1-k)(1+r_i)]^n \tag{10}$$

*Then, the upper bound of $R(n)$, has been proved.*
**End of proof.**

From **Theorem 1,** we can present the proposition as follows,

**Proposition 1:**    $R(n) \leq 1$ and $\lim_{n \to +\infty} R^n = 0$, if $\bar{r} \leq k$

**Proof** *Because $\bar{r} \leq k$, therefore*

$$R(n) \leq [(1-k)(1+r_i)]^n \leq (1-k^2)^n \leq 1 \tag{11}$$

because, $k \in (0,1)$, $(1 - k^2) \in (0,1)$, so $R(n) \leq 1$. Further,

$$0 \leq \lim_{n \to +\infty} R^n \leq \lim_{n \to +\infty} (1 - k^2)^n = 0 \tag{12}$$

**End of proof**

Obviously, the Proposition 1 reveals the fact that when the number of trades $n$ increases to infinity, the cumulative returns will converge to 0, which means investors will lose all of invested capital.

# III. Dataset and Methodology

## A. Data Materials

In this paper, the profitability of proposed technical trading rules is assessed on four important financial market indices, which are Dow Jones Industrial Average (DJIA), FTSE 100 Index (FTSE), Nikkei 225 (N225) and SSE Composite Index (SCI), respectively. And we download these daily indices from Yahoo Finance, a popular publicly available source of data. Further, we show the temporal evolution of these stock indices in Fig 1.

- DJIA, from September, 12th 1988 to September, 09th 2018, for a total of 7562 days;
- FTSE, from September, 12th 1988 to November, 26th 2018, for a total of 7663 days;
- N225, from September, 13th 1988 to November, 26th 2018, for a total of 7408 days;
- SCI, from January, 04th 2000 to November, 26th 2018, for a total of 5464 days.

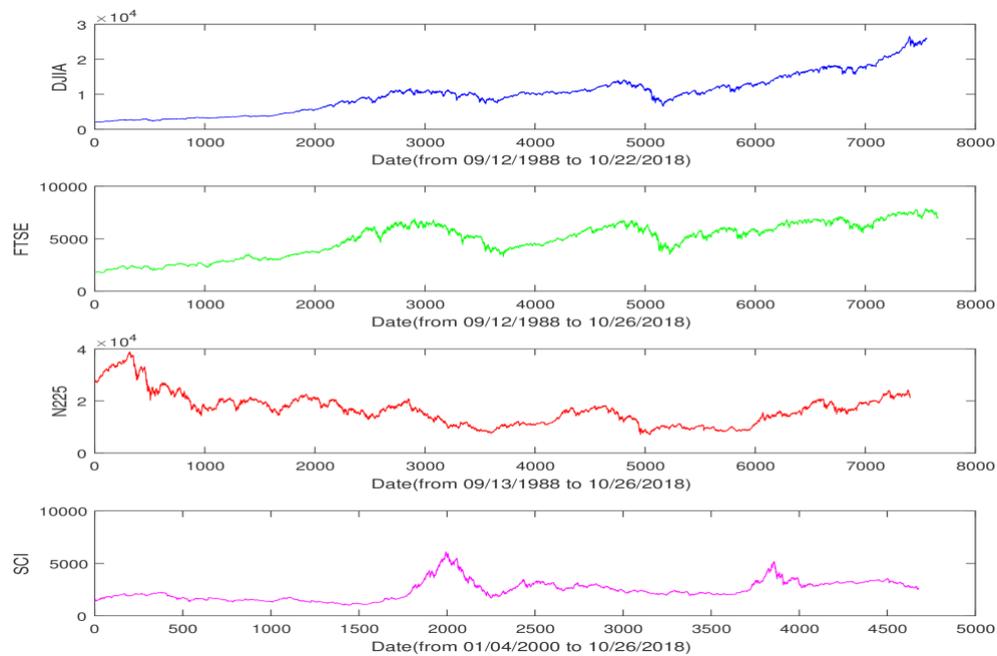

**Fig 1. T**emporal evolution of four stock indices, i.e.**,** DJIA, FTSE, N225 and SCI indices.

## B. Technical Trading Rules

As we all know, technical trading rules are widely used in stock market investment. In this part, we will introduce 13 the most popular technical trading rules, which are simple moving average (SMA), exponential moving average (EMA), momentum (MOM), stochastic oscillator (KD), moving average convergence divergence(MACD), relative strength index(RSI), psychology (PSY), commodity channel index (CCI), moving average (MA), BIAS, rate of change(ROC), directional movement index (DMI) and random trading strategy (RND), respectively. As the simplest trading strategy [8], the random trading strategy makes trade decision (buy or sell) at time t completely at random (t follows a uniform distribution with a mean of 15). The other 12 technical trading rules and their formula are presented in Table 1. Notably, the parameter setting of the 12 trading rules is according to [24].

Table 1. Details of technical trading rules.

| Oscillator | Formula | Parameter | Technical rules Buy | Technical rules Sell |
|---|---|---|---|---|
| SMA | $\text{SMA}(n) = \frac{1}{n}\sum_{i=0}^{n} C_{t-i}$ | n=20 | SMA↗C | SMA↘C |
| EMA | $\text{EMA}(n) = \left[(C_t - EMA_{t-1}) \times \frac{1}{n+1}\right] + EMA_{t-1}$ | n=5, m20 | EMA5 ↗ EMA20 | EMA5 ↘ EMA20 |
| MOM | $\text{MOM}(n) = \frac{C_t}{C_{t-n}} \times 100$ | n=10 | MOM ↗ 0 | MOM ↘ 0 |
| KD | $K(n) = \frac{C_t - L_{n,min}}{H_{n,max} - L_{n,min}} \times 100$ <br> $D(n) = \sum_{i=0}^{n-1} K_{t-i}/n$ | n=12, m=12 | K↗D & D<20 | K↘D & D>80 |
| MACD | $\text{MACD}(m,n) = \text{EMA}(n) - \text{EMA}(m)$ | n=12, m=26 | MACD↗0 | MACD↘0 |
| RSI | $\text{RSI}(n) = 100 - \frac{100}{1 + RS(n)}$ <br> $RS(n) = \sum_{i=0}^{n-1} Up_{t-i} / \sum_{i=0}^{n-1} Down_{t-i}$ | n=14 | RSI↗30 | RSI↘70 |
| PSY | $\text{PSY}(n) = \frac{number\ of\ up\ trend}{n}$ | n=10 | PSY ↗ 25% | PSY ↘ 75% |
| CCI | $\text{CCI}(n) = \frac{M - M(n)}{d(n) \times 0.015}$ <br> $M = \frac{H + L + C}{3}$ <br> $d(n) = \frac{1}{n}\sum_{i=0}^{n-1}|M_{t-i} - \bar{M}_t(n)|$ | n=9 | CCI ↗ -100 | CCI ↘ 100 |
| MA | $\text{MA}(n) = \frac{1}{n}\sum_{i=0}^{n} C_{t-i}$ | n=5, m=20 | MA5 ↗ MA20 | MA5 ↘ MA20 |
| BIAS | $\text{BIAS}(n) = \frac{C_t - MA(n)}{MA(n)}$ | n=10 | BIAS ↗ -4.5% | BIAS ↘ 5% |

| ROC | $\text{ROC}(n) = (\frac{C_t}{C_{t-n}} - 1) \times 100$ | n=13 | ROC↗0 | ROC↘0 |
| --- | --- | --- | --- | --- |
| DMI | $+\text{DI} = \frac{+DM}{TR} \times 100 \ ; \ -\text{DI} = \frac{-DM}{TR} \times 100$ | n=14 | +DI↗-DI | +DI↘-DI |
| | $+DM = \max(H_t - H_{t-1}, 0)$ | | | |
| | $-DM = \max(L_t - L_{t-1}, 0)$ | | | |
| | $TR = \max(H_t - L_t, H_t - C_{t-1}, L_t - C_{t-1})$ | | | |

[a] C, L, H, and Up(Down) are the close price, low price, high price and upward(downward) price change, respectively; ↗,↘ mean upwards/downwards cross;

[b] $y_{n,max} = \max(y_t, \dots, y_{t-n+1})$, $y_{n,mim} = \min(y_t, \dots, y_{t-n+1})$, $\bar{y}(n) = \frac{1}{n}\sum_{i=0}^{n-1} y_{t-i}$.

## C. Bootstrap Methodology

In order to reduce the influence of "luck" and make results more convincing, we adopt bootstrap methodology [17] in experiments. The main steps of the bootstrap methodology in this paper are described as follows.

1. Resample: For each experiment, randomly choose entering and exiting points, forming a test period $[Enter\_point_i, Exit\_point_i]$;
2. Generate return series and compute cumulative return: In each test period, use a certain trading rule to make trades and obtain the corresponding return series, then compute cumulative return $R_i$ by Eq.3;
3. Repeat step 1 and step 2: Repeat the above two steps for M times, then get the estimation of $\bar{R}$, $\bar{R} = \frac{1}{M}\sum_{i=0}^{M} R_i$. In this paper, we take M as 1000.

# IV. Experiments and Analyses

To investigate the profitability of used technical trading rules, we first analyze the cumulative return R(n) on the four indices from the perspective of the mean of return rate series ($\bar{r}$), number of trades (n) and transaction cost rate (k). Furthermore, we also conduct a number of experiments to evaluate the profitability of the proposed technical trading rules by comparing with the market and random trading strategy. It is worthy to mention that all the results are generated by bootstrap methodology.

## A. Analyses on Cumulative Return

To testify the effectiveness of proposed upper bound, we choose DJIA as an example to illustrate the influence of the number of trades on cumulative return and its upper bound (here use the random trading strategy, i.e., Randomly Buy and Hold, noted as R* in Fig2). Fig 2 shows the results. Easily find that regardless of transaction cost rate k, Eq. 5 holds. In addition, when k = 0.007, it satisfies the condition that $\bar{r} \leq k$ (see Table 2, $\bar{r} = 0.0048$), the cumulative return and its upper bound show a clearly downward tendency, which makes the validity of Proposition 1 to some extent.

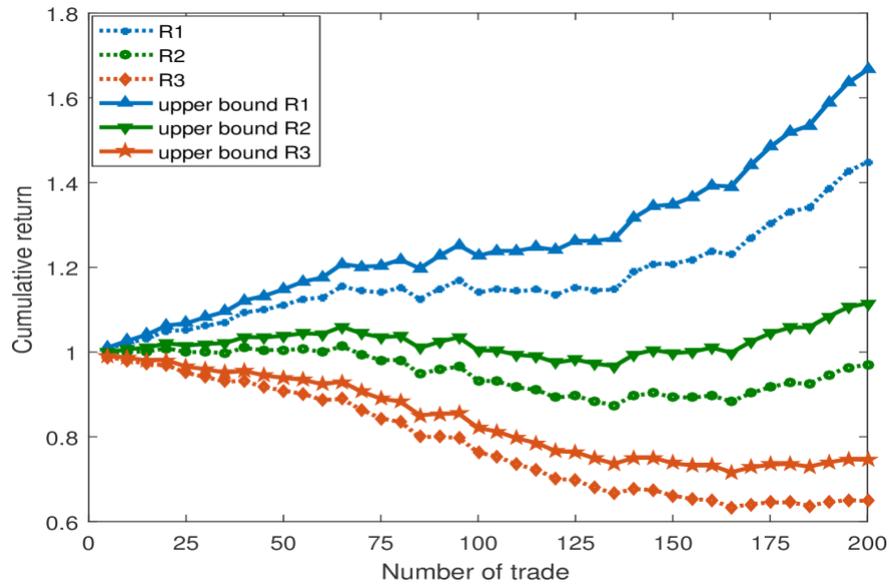

**Fig 2.** Validation of the upper bound of cumulative return in DJIA. Here R1, R2 and R3 are the cumulative return when $k$ takes 0.003, 0.005 and 0.007 respectively. As visible, the validity of upper bound is verified, here we have no consideration of transaction cost rate $k$.

We also find the $\bar{r}$ have a significant impact on the upper bound of cumulative return (see Eq. 5). Therefore, it is necessary to investigate the $\bar{r}$ of technical trading rules in different stock indices. The results are presented in Table 2. To some extent, the $\bar{r}$ can reflect the potential profitability of a certain technical trading rule in some stock indices, such as RSI may have a better performance in DJIA and FTSE than N225 and SCI, according to $\bar{r}$ in Table 2.

**Table 2.** The $\bar{r}$ of used technical trading rules in stock indices

| Trading rules | Stock indices | | | |
|---|---|---|---|---|
| | DJIA | FTSE | N225 | SCI |
| **BIAS** | -0.0044 | -0.0320 | -0.0262 | 0.0149 |
| **CCI** | 0.0046 | 0.0052 | 0.0002 | -0.0021 |
| **DMI** | 0.0025 | 0.0028 | 0.0035 | 0.0059 |
| **SMA** | 0.0024 | 0.0008 | 0.0005 | 0.0195 |
| **EMA** | 0.0040 | 0.0014 | 0.0005 | 0.0281 |
| **KD** | 0.0939 | 0.0853 | 0.0102 | 0.0608 |
| **MA** | 0.0034 | 0.0014 | 0.0023 | 0.0167 |
| **MACD** | 0.0065 | 0.0023 | 0.0059 | 0.0251 |
| **MOM** | 0.0018 | -0.0007 | 0.0006 | 0.0107 |
| **PSY** | 0.0337 | 0.0295 | 0.0230 | 0.0037 |
| **RND** | 0.0048 | 0.0028 | 0.0004 | 0.0048 |
| **ROC** | 0.0038 | 0.0016 | 0.0005 | 0.0149 |
| **RSI** | 0.0610 | 0.0220 | -0.0369 | -0.0363 |

Furthermore, the influence of transaction costs is investigated. Transaction cost is an important factor that affects the cumulative return and always consists of two major components: explicit costs and implicit costs. The former is the direct costs of trading, such as broker commissions and taxes, while the latter involves in indirect costs such as the influence of the trade price and the opportunity cost of failing to execute the order, which is difficult to measure [16]. By the way, it should be noted that in many markets, especially in the emerging markets, the implicit costs are even higher than the explicit costs [28]. In fact, the transaction cost can affect the cumulative return of trading strategies drastically [16]. In simplicity, we only take explicit costs into consideration and approximately measure transaction costs by transaction cost rate $k$ in this paper. For the sake of space, here we take the ROC, CCI, RND, and MACD strategies as examples to show the influence of $k$ on $R(n)$ in different stock indices, and set the value of $k$ range from 0.001 to 0.01. We display the results in Fig 3. Undoubtedly, the cumulative return show a clear downward tendency when the $k$ increase gradually, this is in accordance with common sense. We also notice that the ROC and MACD achieve higher profits in SCI, while CCI and RND more succeed in FTSE and DJIA, respectively. So we conclude different technical trading rules often have different performance on specific stock indices.

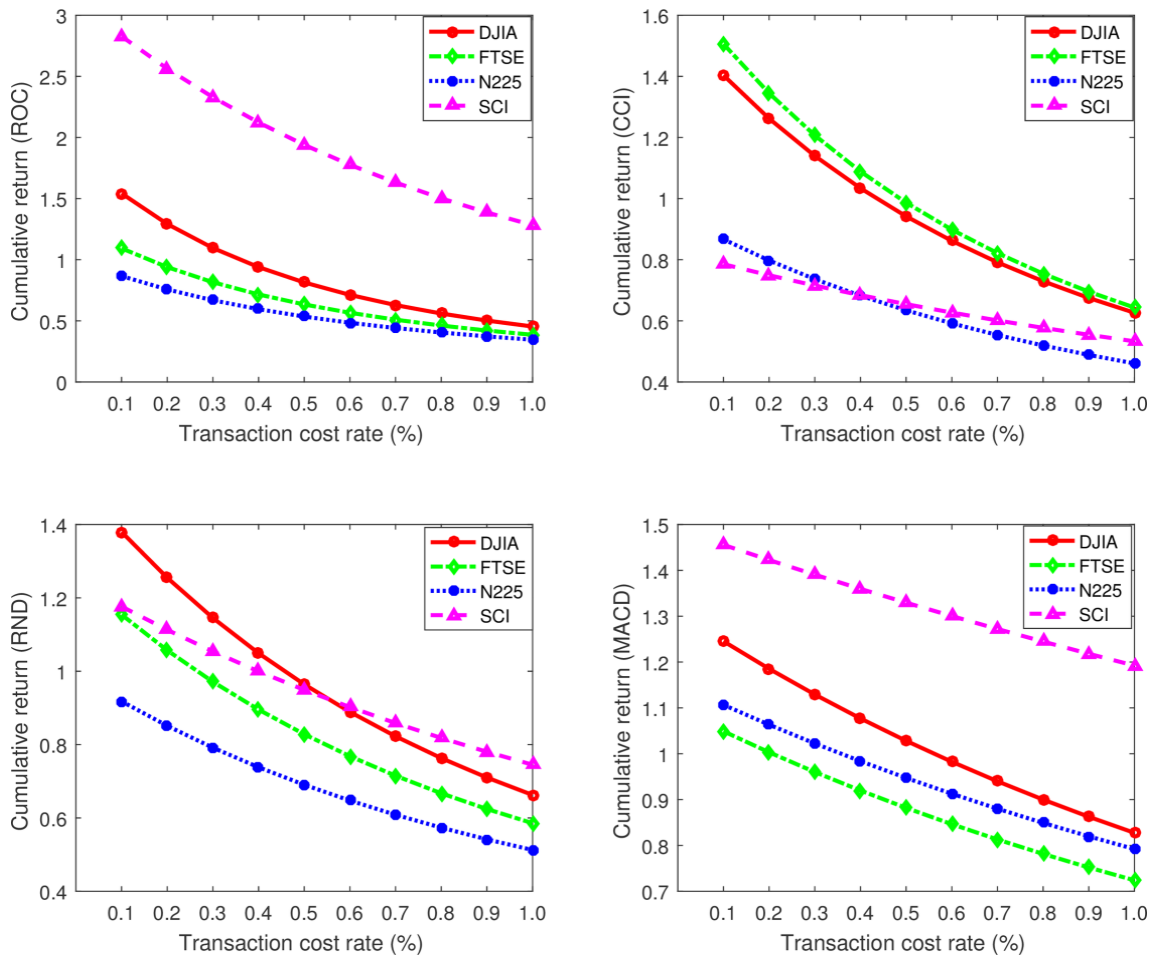

**Fig 3.** The influence of $k$ on cumulative return in stock indices, i.e., the results of ROC, RND, CCI and MACD strategies, respectively.

In addition, we also investigate the influence of the number of trades n on $R(n)$. Here we take DJIA as an example and exhibit results in Fig 4. From Table 2, we have known the $\bar{r}$ of the ROC, CCI, RND and MACD in DJIA are 0.0038, 0.0046, 0.0048, 0.0065, respectively. When $k$ takes 0.007, satisfied $k > \bar{r}$, therefore according to Proposition 1, the $R(n)$ ought to show a approximately downward tendency when n increases gradually. Apparently, Fig 4 (k = 0.007) supports our inference. As for when n is small, there exists $R(n) > 1$, we believe it mainly caused by the "luck". While with n increases, the influence of the "luck" will diminish gradually. Other subgraphs in Fig 4 deliver that if $k < \bar{r}$, it has a large probability that $R(n)$ will increase with the n increases. Therefore, the relationship between k and $\bar{r}$ is a crucial factor in evaluating profitability of trading strategy in investment.

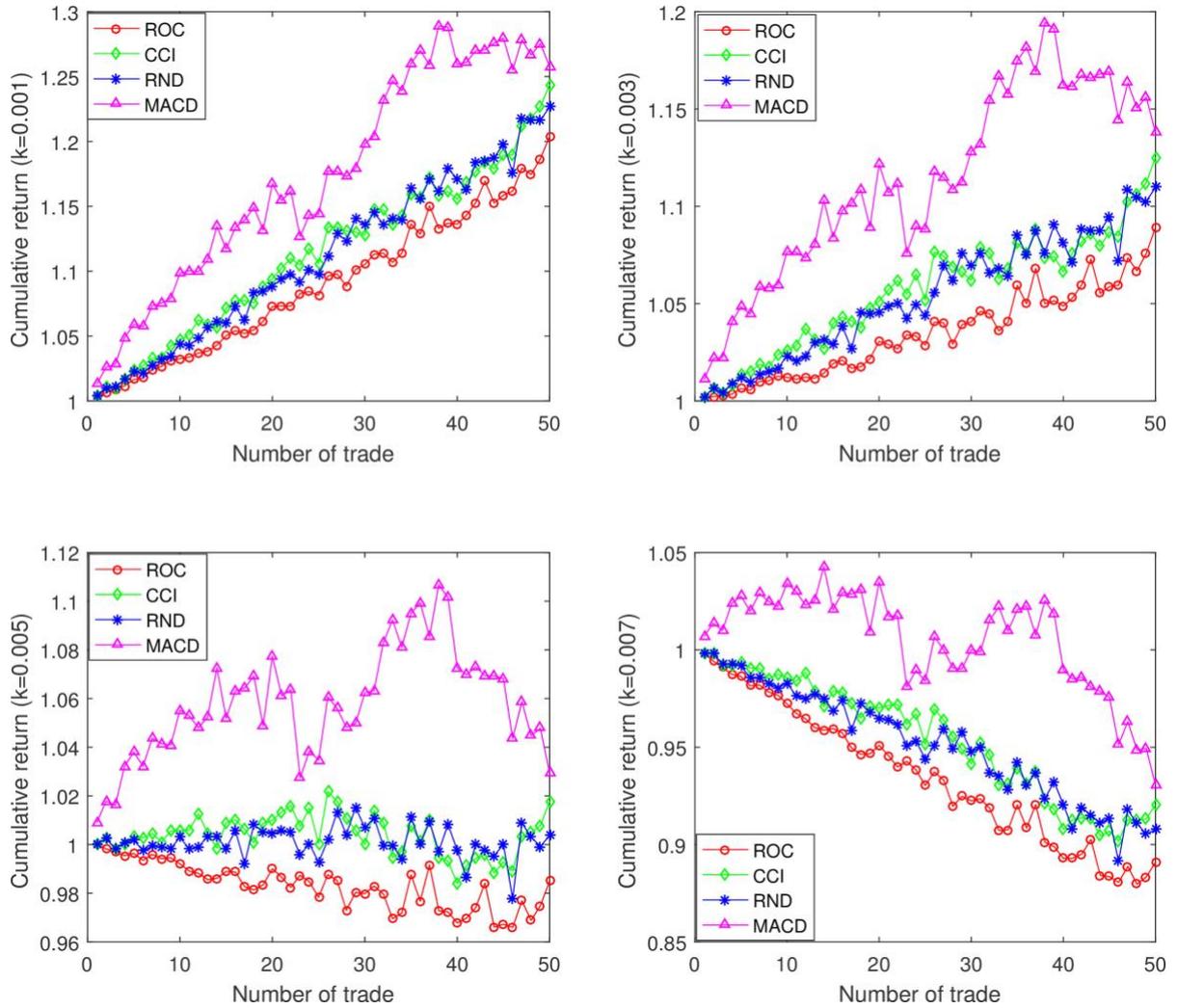

**Fig 4.** The influence of the number of trades n on cumulative return in DJIA. From top to bottom, from left to right, k takes 0.001, 0.005, 0.003 and 0.007 respectively. See text for further analysis.

## B. Profitability Evaluation

We have introduced several common technical trading rules and have obtained their $\bar{r}$ (see Table 2) in the former analysis. Besides, we also conclude that when $k > \bar{r}$, the technical trading rules are not enabled to gain profits. Although some technical trading rules can achieve great $\bar{r}$ in some stock indices, the profitability of these trading strategies is still not convincing. Therefore, this part aims to testify and assess their profitability. Before conducting experiments, we first introduce the compound annual growth rate (CAGR), which is utilized to evaluate the profitability of used technical trading rules, which is

calculated as follows:

$$R_{CAGR} = \sqrt[Y]{R} - 1 \qquad (13)$$

Here, $R$ is cumulative return (see Eq. 3), and $Y$ is number of years.

In practice, we also compute the CAGR of the market value of a stock index for comparison, denoted as CMV. Former analysis has indicated that the relationship between $k$ and $\bar{r}$ is of great importance in assessing profitability. However, the value of $k$ tends to be correlated with the stock market [29]. For example, in the Chinese stock market, the transaction costs mainly consist of stamp duty (0.1% of the turnover), commission (no more than 0.3% of the turnover) and transfer fees (only charge in the Shanghai market, which is 0.006% of the denomination). Therefore, to simplify the process of estimating transaction costs, we take $k = 0.003$, which is acceptable in Chinese stock markets. Then Table 3 and Fig 5 exhibit the CAGR of proposed technical trading rules. Easily find none of these technical trading rules achieve satisfactory results in N225, this may be caused by downwards tendency of N225 (see Fig 1). In addition, just a few trading strategies get a better CAGR than CMV, such as SMA, EMA, and MOM strategies in SCI. In this respect, some technical trading rules show a better performance on emerging market (SCI) than developed markets (DJIA, FTSE, N225). In terms of profitability, most of the technical trading rules fail to beat the market and are even less effective than the random trading strategy (RND) in Fig 5. In addition, most of the technical trading rules have different performance on different stock indices, which further suggests that the profitability of technical trading rules is unstable.

**Table 3. The CAGR of used technical trading rules in stock indices**

|  |  | DJIA | FTSE | N225 | SCI |
|---|---|---|---|---|---|
| **CMV**[a] |  | 0.0768 | 0.0405 | -0.0122 | 0.0499 |
| **Trading rules** | **BIAS** | -0.0052 | -0.0159 | -0.0339 | 0.0002 |
|  | **CCI** | 0.0138 | 0.0201 | -0.0362 | -0.0566 |
|  | **DMI** | -0.0141 | -0.0100 | 0.0053 | 0.0543 |
|  | **SMA** | -0.0157 | -0.0372 | -0.0415 | 0.1328 |
|  | **EMA** | 0.0006 | -0.0184 | -0.0305 | 0.1107 |
|  | **KD** | 0.0571 | 0.0388 | 0.0016 | 0.0396 |
|  | **MA** | -0.0045 | -0.0191 | -0.0146 | 0.0554 |
|  | **MACD** | 0.0085 | -0.0082 | -0.0017 | 0.0414 |
|  | **MOM** | -0.0279 | -0.0569 | -0.0459 | 0.0815 |
|  | **PSY** | 0.0438 | 0.0372 | 0.0254 | 0.0024 |
|  | **RND** | 0.0112 | -0.0060 | -0.0300 | -0.0025 |
|  | **ROC** | 0.0050 | -0.0253 | -0.0454 | 0.1014 |
|  | **RSI** | 0.0466 | 0.0137 | -0.0359 | -0.0459 |

[a] Note: CMV denotes the CAGR of the market value of a stock index in the testing periods.

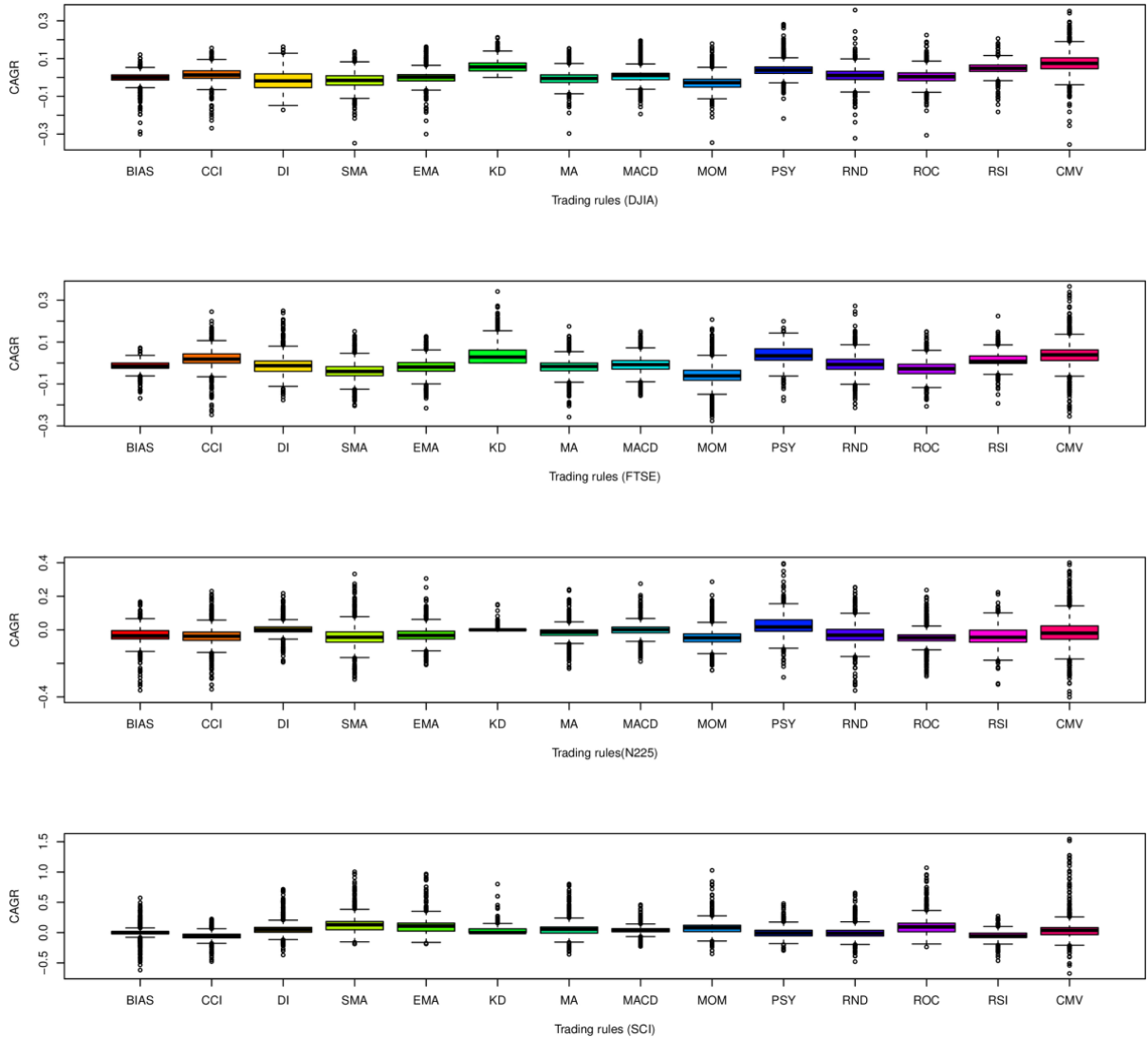

**Fig 5.** The box plot of CAGR of technical trading strategies in four stock indices (here $k = 0.003$). From the top to the bottom, we exhibit the results of DJIA, FTSE, N225 and SCI, respectively.

# V. Conclusion

This paper first presents and proves the upper bound of cumulative return, and further disclose that if the mean of return rate in trading series $\bar{r}$ is less than transaction cost rate $k$, then the more trades, the more losses, and the cumulative return would even converge to zero when the number of trades becomes infinity. Then, we conduct a number of experiments on SCI, DJIA, N225, and FTSE in order to evaluate the profitability of these technical trading rules related to the proposed upper bound of cumulative return. The results show that the used technical trading rules cannot provide stable and satisfactory profits. In terms of profitability, most of the trading strategies fail to beat the market and are even less effective than random ones. It is worth mentioning that we do not deny that technical trading rules with certain parameters can make great profits in some indexes, but the selection of parameters for a stock index is not easy for unprofessional investors. Therefore, we advise unprofessional individual investors who have less investment experience to stay away from stock markets or to seek guidance from professional institutional investors, in case of unnecessary loss, if the average return $\bar{r}$ is certainly more than one in the trading horizon.

We expect the upper bound of cumulative return can provide a new perspective for investors to assess the profitability of a trading strategy. In addition, this paper reveals the fact that when $\bar{r}$ less than $k$, the increase of trade times cannot improve cumulative return. This finding has an important implication for high-frequency trading. In future work, we should pay more attention to find new strategies to reduce unnecessary trades.

# Reference

[1]. Barak S, Dahooie JH, Tich`y T. Wrapper ANFIS-ICA method to do stock market timing and feature selection on the basis of Japanese Candlestick. Expert Systems with Applications. 2015;42(23):9221–9235.

[2]. Arvalo R, Garca J, Guijarro F, Peris A. A dynamic trading rule based on filtered flag pattern recognition for stock market price forecasting. Expert Systems with Applications. 2017;81(C):177–192.

[3]. Hu Y, Liu K, Zhang X, Su L, Ngai E, Liu M. Application of evolutionary computation for rule discovery in stock algorithmic trading: A literature review. Applied Soft Computing. 2015;36:534–551.

[4]. Fernandez E, Navarro J, Solares E, Coello CC. A novel approach to select the best portfolio considering the preferences of the decision maker. Swarm and Evolutionary Computation. 2019;46:140–153.

[5]. Wang Q, Xu W, Zheng H. Combining the wisdom of crowds and technical analysis for financial market prediction using deep random subspace ensembles. Neurocomputing. 2018;299:51–61.

[6]. Macedo LL, Godinho P, Alves MJ. Mean-semivariance portfolio optimization with multiobjective evolutionary algorithms and technical analysis rules. Expert Systems with Applications. 2017;79:33–43.

[7]. Yan Z, Gong R. Game Behavior between Institutional and Individual Investors in Chinese Stock Market. In: Advances in Computer Science and Engineering. Springer; 2012. p. 455–462.

[8]. Biondo AE, Pluchino A, Rapisarda A, Helbing D. Are Random Trading Strategies More Successful than Technical Ones. PLOS ONE. 2013;8(7).

[9]. Fama EF, Blume ME. Filter rules and stock-market trading. The Journal of Business. 1966;39(1):226–241.

[10]. Levy RA. Random walks: Reality or myth. Financial Analysts Journal. 1967;23(6):69–77.

[11]. Fang J, Jacobsen B, Qin Y. Predictability of the Simple Technical Trading Rules: An Out-of-Sample Test. Review of Financial Economics. 2014;23(1):30–45.

[12]. Brock W, Lakonishok J, Lebaron B. Simple Technical Trading Rules and the Stochastic Properties of Stock Returns. Journal of Finance. 1992;47(5):1731–1764.

[13]. Thenmozhi M, Chand GS. Forecasting stock returns based on information transmission across global markets using support vector machines. Neural Computing and Applications. 2016;27(4):805–824.

[14]. Shynkevich Y, Mcginnity TM, Coleman SA, Belatreche A, Li Y. Forecasting price movements using technical indicators: Investigating the impact of varying input window length. Neurocomputing. 2017;264:71–88.

[15]. Patel J, Shah S, Thakkar P, Kotecha K. Predicting stock and stock price index movement using Trend Deterministic Data Preparation and machine learning techniques. Expert Systems With Applications. 2015;42(1):259–268.

[16]. Bajgrowicz PG, Scaillet O. Technical Trading Revisited: False Discoveries, Persistence Tests, and

[17]. Zhu H, Jiang ZQ, Li SP, Zhou WX. Profitability of simple technical trading rules of Chinese stock exchange indexes. Physica A: Statistical Mechanics and its Applications. 2015;439:75–84.

[16 continued]. Transaction Costs. Journal of Financial Economics. 2012;106(3):473–491.

[18]. Liang Y, Yang G, Huang JP. Progress in physical properties of Chinese stock markets. Frontiers of Physics. 2013;8(4):438–450.

[19]. Ruan Q, Yang H, Lv D, Zhang S. Cross-correlations between individual investor sentiment and Chinese stock market return: New perspective based on MF-DCCA. Physica A-statistical Mechanics and Its Applications.2018;503:243–256.

[20]. Wan YL, Xie WJ, Gu GF, Jiang ZQ, Chen W, Xiong X, et al. Statistical properties and pre-hit dynamics of price limit hits in the Chinese stock markets. PloS one. 2015;10(4):e0120312.

[21]. Li H, Zheng D, Chen J. Effectiveness, cause and impact of price limit: Evidence from China's cross-listed stocks. Journal of International Financial Markets, Institutions and Money. 2014;29:217–241.

[22]. Chen G, Kim KA, Nofsinger JR, Rui OM. Trading Performance, Disposition Effect, Overconfidence, Representativeness Bias, and Experience of Emerging Market Investors. Journal of Behavioral Decision Making. 2007;20(4):425–451.

[23]. Dragomir SS, Ionescu NM. Some converse of Jensen's inequality and applications. Rev Anal Numér Théor Approx. 1994;23(1):71–78.

[24]. Chen CH, Chen YH, Lin JCW, Wu ME. An effective approach for obtaining a group trading strategy portfolio using grouping genetic algorithm. IEEE Access. 2019;7:7313–7325.

[25]. Mńg JCP. Dynamically Adjustable Moving Average (AMA´) technical analysis indicator to forecast Asian Tigers´futures markets. Physica A: Statistical Mechanics and its Applications. 2018;509:336–345.

[26]. Kim Y, Enke D. Developing a rule change trading system for the futures market using rough set analysis. Expert Systems with Applications. 2016;59:165–173.

[27]. Hesterberg T, Moore DS, Monaghan S, Clipson A, Epstein R. Bootstrap methods and permutation tests. Introduction to the Practice of Statistics. 2005;5:1–70.

[28]. Moreno D, Olmeda I. Is the predictability of emerging and developed stock markets really exploitable? European Journal of Operational Research. 2007;182(1):436–454.

[29]. Domowitz I, Glen J, Madhavan A. Liquidity, volatility and equity trading costs across countries and over time. International Finance. 2001;4(2):221–255.

## Declarations